\vsize=24.truecm \hsize=16truecm
\baselineskip=0.5truecm \parindent=0truecm
\parskip=0.2cm \hfuzz=1truecm
\font\scap=cmcsc10
\font\small=cmr8

\font\tenmsb=msbm10
\font\sevenmsb=msbm7
\font\fivemsb=msbm5
\newfam\msbfam
\textfont\msbfam=\tenmsb
\scriptfont\msbfam=\sevenmsb
\scriptscriptfont\msbfam=\fivemsb
\def\Bbb#1{{\fam\msbfam\relax#1}}

\newcount\eqnumber
\eqnumber=1
\def\neweq{{\rm{(\the\eqnumber)}}\global\advance\eqnumber by 1}
\def\eqdef#1{\eqno\xdef#1{\the\eqnumber}\neweq}
\def\newaeq{{\rm{(\the\eqnumber a)}}\global\advance\eqnumber by 1}
\def\eqdaf#1{\eqno\xdef#1{\the\eqnumber}\newaeq}
\def\eqdisp#1{\xdef#1{\the\eqnumber}\neweq}
\def\eqdasp#1{\xdef#1{\the\eqnumber}\newaeq}

\newcount\refnumber
\refnumber=1
\def\newref{{\the\refnumber}\global\advance\refnumber by 1}
\def\refdef#1{{\xdef#1{\the\refnumber}}\newref}

\centerline{\bf Singularity patterns and dynamical degrees}
\bigskip\bigskip
\hskip.55cm{\sl Graduate School of Mathematical Sciences, the University of Tokyo}\hfill{MASE Takafumi}~~
\smallskip
\hskip.5cm{\sl Graduate School of Mathematical Sciences, the University of Tokyo}\hfill{WILLOX Ralph}~~
\smallskip
\hskip.5cm{\sl IMNC, Universit\'e Paris VII \& XI, CNRS, UMR 8165, Orsay, France}\hfill {RAMANI Alfred}~~
\smallskip
\hskip.5cm{\sl IMNC, Universit\'e Paris VII \& XI, CNRS, UMR 8165, Orsay, France}\hfill {GRAMMATICOS Basil}~~
\bigskip
{\small We explain on a selection of mappings how the method introduced by Halburd and our simplified variant thereof, the so-called express method, can be used to calculate the dynamical degree of second-order rational mappings from nothing more than their singularity structure.}
\bigskip\bigskip
1. {\scap Introduction}
\medskip
Ever since discrete integrable systems came into the limelight, in the past two decades, the study of singularities has become ineluctable.   The reason for this is that integrability is inextricably related to a special singularity structure. Before proceeding further, let us clarify the notion of `singularity'. A singularity appears in a discrete system when at some iteration a degree of freedom is lost. Since in this paper we shall deal exclusively with second-order rational mappings (although the extension of the approach to higher order systems appears possible), a loss of a degree of freedom means that the value of the dependent variable at point $n+1$, is independent of its value at point $n-1$. The special singularity structure we are referring to is that of confined singularities. The term `confinement' [\refdef\sincon] describes a situation where a singularity that appears at some iteration, disappears after a few more iteration steps when the lost degree of freedom is recovered by lifting the indeterminacy that arises at that iteration. The relation of singularity confinement to integrability was based on the observation that systems integrable through the application of the isomonodromy approach do possess the confinement property (with no known counterexample).

Another property intimately related to integrability is that of degree growth, a property akin to that of complexity introduced by Arnold [\refdef\arnold].  According to Veselov [\refdef\veselov], ``integrability has an essential correlation with the weak growth of certain characteristics'' and the best way to make this statement more precise is to consider the dynamical degree of the mapping. The latter is obtained from the degrees $d_n$ of the iterates of some initial condition and is defined as
$$\lambda=\lim_{n\to\infty} d_n^{{1/n}}.$$
Integrable mappings have a dynamical degree equal to 1, while a dynamical degree greater than 1 indicates nonintegrability. The degree growth of a rational mapping is closely related to its singularity structure and to the confinement property. The low growth in integrable mappings results from the fact that algebraic simplifications occur during the iteration of the mapping. They have as a result that the degree grows polynomially with the number of iterations, while in the absence of simplifications, the degree growth is exponential. 

How does one go about calculating the dynamical degree of a given mapping? The traditional way is the heuristic one: one establishes the behaviour of $d_n$ based on the explicit computation of a sufficient number of iterates [\refdef\bellon]. For example by introducing initial conditions $x_0$ and $x_1=z$, where $x_0$ is a completely generic complex number, i.e. not supposed to satisfy any particular algebraic relation, and  $z\in{\Bbb C}\cup\{\infty\}$. One then calculates the degree $d_n$ in $z$ of the $n$th iterate of the mapping and by computing the limit of the ratio of two successive $d_n$'s one can estimate the dynamical degree. A rigorous approach is also possible [\refdef\favre], at least for autonomous mappings. In the confining case the latter consists in performing the regularisation of the mapping, through a (finite) sequence of blow-ups, to an automorphism of the surface obtained from these blow-ups. The dynamical degree of the mapping is then obtained as the largest eigenvalue of this automorphism [\favre], [\refdef\take]. For nonconfining mappings the approach is less systematic, but the general theory  [\favre] tells us that, generically, the dynamical degree will be greater than 1 (unless the mapping is a linearisable one, in which case $\lambda=1$). 

However a third approach is also possible, thanks to the pioneering work of Halburd [\refdef\rod]. One starts from the observation that the degree of the $n$th iterate $f_n(z)$ of the mapping is equal to the number of preimages of some arbitrary value $w\in{\Bbb C}\cup\{\infty\}$ for that function. This is tantamount to counting (with the appropriate multiplicity) the number of solutions, in $z$, of the equation $f_n(z)=w$.
Halburd's innovative method [\rod] consists in computing the degree, not for just any arbitrary value of  $w$, but for special values which appear in the singularity patterns of the mapping. The counting of preimages is then performed based on information that can be obtained from the singularity analysis of the mapping. The precise workings of Halburd's method will be explained in the sections that follow.
\bigskip
2. {\scap A collection of results}
\medskip
In order to be able to apply Halburd's method [\rod] one must first establish the precise singularity structure of the mapping one is studying. This means that {\sl all} singularity patterns must be obtained. As we explained in [\refdef\anticonf], the singularities that appear in rational mappings are not only of confined or unconfined type. Two more singularity types can exist.  In the case of those we have dubbed `cyclic', a pattern keeps repeating for all iterations. On the other hand, those we have called `anticonfined' correspond to patterns in which singular values persist indefinitely in both the forward and backward iteration, with just a finite region of regular values in between. Not all the aforementioned types exist for all mappings, but it is essential for the application of Halburd's method that all existing patterns be obtained.  

On the other hand, a much simpler method has been proposed under the moniker of `express' method, in previous works [\refdef\rodone,\refdef\rodnon] of ours. The advantage of our method is that it uses only the information coming from the confined and/or unconfined patterns in order to obtain the dynamical degree of a given mapping. When the mapping is nonintegrable, i.e. when the dynamical degree is greater than 1, the express method is largely sufficient. And in the case of integrable mappings, the method gives correctly the value of the dynamical degree, namely 1. However if one wishes to obtain the exact value of the degree growth, then one must go back to the full method of Halburd and take into account the cyclic and/or anticonfined singularities as well.
\medskip
{\sl The discrete Painlev\'e I}
\smallskip
We start with a well-known integrable case, namely the discrete Painlev\'e I:
$$x_{n+1}+x_{n-1}={a_n\over x_n}+{1\over x_n^2},\eqdef\zena$$
where $a_n$ satisfies the relation $a_{n+1}-2a_n+a_{n-1}=0$.
It has the confined singularity pattern
$$\{0,\infty^2,0\},$$
where the exponent of $\infty$ must be interpreted as follows: had we introduced a small  quantity $\epsilon$ by assuming that $x_n=\epsilon$, we would have found that $x_{n+1}$ is of the order of $1/\epsilon^2$.

Let us denote by $Z_n$ the number of spontaneous occurrences of 0 at some step $n$. Under the iteration of the mapping, the total number of times the value zero occurs at step $n$, is equal to the sum of $Z_n$ and the number of zeros that result, in the singularity pattern, from a spontaneous occurrence of 0 two steps before, namely $Z_{n-2}$. Thus the degree $d_n$, as obtained from the preimages  of 0, is given by $d_n=Z_n+Z_{n-2}$. Given the form of (\zena), the total number of times $\infty$ occurs at step $n$ is given by the number of spontaneous occurrences of 0 at the previous step (but notice that the multiplicity of $\infty$ is 2), i.e. $2Z_{n-1}$, plus the occurrences of $\infty$ arising from a cyclic pattern:
$$\{x_0,\infty\}.$$
Note that as this cyclic pattern $x_0, \infty, -x_0, \infty, x_0, \infty, \cdots$ does not contain any singularities, it is usually not discussed at all in standard singularity analysis.

The contribution of the above cyclic pattern to the degree, counted as the number of preimages of $\infty$, is $(1-(-1)^n)/2$ since an infinity appears every two steps, and we find $d_n = 2 Z_{n-1} + (1-(-1)^n)/2$.  Equating the expressions for the degree obtained from the preimages of 0 and $\infty$, we obtain the relation
$$Z_n+Z_{n-2}=2Z_{n-1}+{1-(-1)^n\over2}.\eqdef\zdyo$$
The solution of (\zdyo) with initial conditions $Z_{0}=0, Z_{1}=1$ is
$$Z_n={n^2\over4}+{n\over2}+{1-(-1)^n\over8},\eqdef\ztri$$
leading to the expression for the degree
$$d_n={2n^2+1-(-1)^n\over4},\eqdef\ztes$$
in perfect agreement with the calculated sequence, 0, 1, 2, 5, 8, 13, 18, 25, 32, 41, 50,$\cdots$.

On the other hand, the express method consists in completely neglecting the contribution of the cyclic pattern. In this case, instead of (\zdyo), one has the equation
$$Z_n+Z_{n-2}-2Z_{n-1}=0,\eqdef\zpen$$
but the relation of $Z_n$ to the degree can no longer be established. Still, as explained in [\rodone], relation (\zpen) can be used to assess the integrability of (\zena). Indeed, the characteristic equation for (\zpen) is $(\lambda-1)^2=0$ which is consistent with the criterion for integrability, namely the absence of a characteristic root greater than 1. Note that (\zpen) is exactly the same equation as that satisfied by $a_n$. In fact, it is this observation that lies at the heart of the full deautonomisation approach described in [\refdef\redeem].
\medskip
{\sl The Hietarinta \& Viallet mapping}
\smallskip
The Hietarinta-Viallet (H-V) [\refdef\hiv] mapping, 
$$x_{n+1}+x_{n-1}=x_n+{1\over x_n^2},\eqdef\zhex$$
is the best known example of a nonintegrable mapping with confined singularities. Its confined singularity pattern is 
$$\{0,\infty^2,\infty^2,0\},$$
and a cyclic singularity pattern also exists:
$$\{x_0,\infty,\infty\}.$$
The degree $d_n$, obtained from the preimages  of 0, is given by $d_n=Z_n+Z_{n-3}$. From the preimages of $\infty$ we have $2Z_{n-1}+2Z_{n-2}+(2-j^n-j^{2n})/3$, with $j=e^{2i\pi/3}$, where the first two terms come from the confined pattern and the last term from the cyclic one. Identifying these two expressions for the degree, we obtain  the equation 
$$Z_n+Z_{n-3}=2Z_{n-1}+2Z_{n-2}+{2-j^n-j^{2n}\over3}.\eqdef\zhep$$
The solution of (\zhep), with initial conditions $Z_{-1}=Z_{0}=0, Z_{1}=1$, is
$$Z_n={\sqrt{5}\over20}\left(\varphi^{2n+3}+\varphi^{-2n-3}\right)-{1\over3}-{j^n+j^{2n}\over12},\eqdef\zoct$$
where $\varphi$ is the golden mean $\varphi=(1+\sqrt5)/2$. This expression for $Z_n$ is consistent with the degree of the mapping directly obtained by iterating the mapping, namely 0, 1, 3, 8, 23, 61, 160, $\cdots$. 

However, if we only care about the dynamical degree of the mapping, it is simpler to use the express method. In this case we neglect the contribution from the cyclic pattern and keep just the homogeneous part of (\zhep) leading to 
$$Z_n+Z_{n-3}-2Z_{n-1}-2Z_{n-2}=0,\eqdef\zenn$$
the characteristic equation of which is 
$$(\lambda+1)(\lambda^2-3\lambda+1)=0.\eqdef\zdek$$
Its largest root is $\lambda=\varphi^2$, a result which can of course be deduced from expression (\zoct)  but which is obtained here in a far simpler way. Note that equation (\zenn) is exactly the same equation as that obtained from the full deautonomisation method, where it allows to conclude on the nonintegrability of (\zhex), despite the existence of confined singularities.
\medskip
{\sl The Bedford \& Kim nonconfining mapping}
\smallskip
The general Bedford-Kim [\refdef\kim] mapping was studied, using the express method in [\rodone]. Here we shall concentrate on a special case of this mapping which has unconfined singularities,
$$x_{n+1}=c{x_n-1\over x_{n-1}-1},\eqdef\dena$$
where $c$ is taken to be a generic, transcendental, number.
The unconfined singularity pattern is
$$\{1,0,\infty,\infty,-c^2,0,{c\over c^2+1},{c(c^2-c+1)\over c^2+1},\cdots\}.$$
Another pattern does also exist, but, upon closer inspection, it becomes clear that it is nothing but the unconfined pattern for the inverse mapping:
$$\{\cdots,1,1,1,x_0,\infty,\infty,{c^2\over x_0-1},0,{c(1-x_0)\over c^2+1-x_0},\cdots\}.$$
Under the express method we neglect any contribution coming from the second pattern. Based only on the first one and denoting by $U_n$ the spontaneous occurrences of 1, we have that the degree obtained from the preimages of 1 is simply $U_n$. The contribution to the degree of the preimages of $\infty$ from the first pattern is $U_{n-2}+U_{n-3}$. Equating the two, we find the equation
$$U_n-U_{n-2}-U_{n-3}=0,\eqdef\ddyo$$
the characteristic equation of which is
$$\lambda^3-\lambda-1=0.\eqdef\dtri$$
(Note that we would have obtained the same equation, up to an inconsequential factor,  if instead we had considered the contribution of the preimages of 0, i.e.: $U_{n-1}+U_{n-5}$). Equation (\dtri) is a well-known one, defining the so-called plastic constant, which is in fact the smallest Pisot number: 1.324718$\dots\,$. The dynamical degree of (\dena) is thus expected to be the plastic constant, which is in perfect agreement with the calculated sequence of degrees: 0, 1, 1, 1, 2, 2, 3, 4, 5, 7, 9, 12, 16, 21, 28, 37, 49, 65, 86, 114, 151, 200,$\cdots$.
\medskip
{\sl A nonconfining extension of the H-V mapping}
\smallskip
In [\redeem] we have studied various extensions of the H-V mapping using the full deautonomisation method. Here we shall concentrate on one of those mappings:
$$x_{n+1}+x_{n-1}=x_n+{1\over x_n^k},\eqdef\dtes$$
where $k$ is an odd integer greater than 2. For such values of $k$, mapping (\dtes) has an unconfined singularity
$$\{0,\infty^k,\infty^k,0,\infty^k,\infty^k,0,\cdots\},$$
as well as a cyclic one $\{x_0,\infty,\infty\}$, but in the spirit of the express method we shall neglect the effect of the latter and consider only the unconfining one. Denoting by $Z_n$ the spontaneous occurrences of 0, we have for the preimages of 0 
$$Z_n+Z_{n-3}+Z_{n-6}+\cdots\equiv\sum_{\ell=0}^{\infty} Z_{n-3\ell},\eqdef\dpen$$
and using the preimages of $\infty$, we obtain the equation
$$\sum_{\ell=0}^{\infty} Z_{n-3\ell}=k\sum_{\ell=0}^{\infty} (Z_{n-3\ell-1}+Z_{n-3\ell-2}).\eqdef\dhex$$
It should be stressed that the sums that appear in this equation are in fact finite ones, as we take all $Z_m$ with negative indices $m$ to be zero.

In order to obtain the characteristic equation we put $Z_n=\lambda^n$ in (\dhex). If we assume now that this characteristic equation has a root with modulus greater than 1, we can sum the infinite series that appear as we take the limit $n\to\infty$. We thus obtain the relation
$${1\over1-{1\over\lambda^3}}=\left({1\over\lambda}+{1\over\lambda^2}\right){k\over1-{1\over\lambda^3}},\eqdef\dhep$$
or, equivalently, 
$$\lambda^2-k\lambda-k=0,\eqdef\doct$$
the largest root of which is $\lambda=(k+\sqrt{k^2+4k\,}\,)/2$. This result is in perfect agreement with the value of the algebraic entropy that was obtained, rigorously, in [\refdef\kanki]. Note also that this value of the dynamical degree is quite different from that for the nonautonomous, confining, version of (\dtes) we studied in [\redeem].
\medskip
{\sl A linearisable mapping}
\smallskip
In [\refdef\linqrt] we have encountered the mapping
$$x_{n+1}x_{n-1}=x_n^2-1,\eqdef\denn$$
which belongs to the class we have called ``linearisable of the third kind''; Its linearisation was presented in [\linqrt]. The degree of mapping (\denn) grows linearly: $d_0=0, d_1=1$ and $d_n=2(n-1)$ for $n>1$. The mapping has two confined singularity patterns
$$\{\pm1,0,\mp1\},$$
and an anticonfined one
$$\{\cdots,\infty^3,\infty^2,\infty,x_0,0,-1/x_0, \infty,\infty^2,\infty^3,\cdots\}.$$
We denote by $U_n, M_n$ the number of spontaneous occurrences of the values $+1, -1$  in the iteration. Since $+1$ and $-1$ obviously play the same role we have $U_n=M_n$ and, thus, find for the degree coming from the preimages of 1
$$U_n+M_{n-2}\equiv U_n+U_{n-2},\eqdef\ddek$$
and, from the preimages of 0, 
$$U_{n-1}+M_{n-1}+\delta_{n1}\equiv2U_{n-1}+\delta_{n1},\eqdef\vena$$
where the Kronecker $\delta_{n1}$ comes from the single appearance of a 0 in the anticonfined pattern. Equating the two expressions we find the equation
$$U_n+U_{n-2}=2U_{n-1}+\delta_{n1}.\eqdef\vdyo$$
Had we decided to simply work with the express method, we would have neglected the $\delta_{n1}$ term in this equation, readily obtaining a dynamical degree equal to 1 from the ensuing homogeneous equation. If we however keep the full expression (\vdyo) we find that $U_n=n$ (under the constraint $U_n=0$ when $n<0$). As a consequence, we obtain exactly the degree $d_n$ we computed empirically.
\medskip
{\sl The mapping of Tsuda et al.}
\smallskip
We shall conclude this selection of examples with a mapping for which, contrary to the previous case, an anticonfined singularity pattern does play an important role:
$$x_{n+1}=x_{n-1}\left(x_n-{1\over x_n}\right),\eqdef\vtri$$
which was introduced in [\refdef\tsuda]. This mapping is nonintegrable despite it having  two confined singularities
$$\{\pm1,0,\infty,\mp1\}.$$
However, it also possesses the anticonfined singularity pattern
$$\{\cdots,0^8,0^5,0^3,0^2,0,0,x,0,\infty,x', \infty,\infty,\infty^2,\infty^3,\infty^5,\infty^8,\cdots\},$$
in which the exponents clearly form a Fibonacci sequence. Due to the exponential growth of the orders of the singularity in this pattern, one surmises that the dynamical degree should be minimally equal to that of the Fibonacci sequence. Calculation of the degree 0, 1, 2, 4, 8, 14, 24, 40, 66, 108, 176, 286, 464, 752, 1218, $\dots$ shows indeed exponential growth, the empirical dynamical degree that can be deduced from this sequence being equal to the golden mean $\varphi=(1+\sqrt 5)/2$. 

Denoting by $U_n$ the number of spontaneous occurrences of the value 1 in the iteration of the mapping ($+1$ and $-1$ clearly playing the same role), we find that the degree at iterate $n$, calculated as the number of preimages of the value 1, is given by
$$U_n+U_{n-3}.\eqdef\vtes$$ 
Similarly, the degree calculated as the number of preimages of 0 is given by
$$2U_{n-1}+\delta_{n1},\eqdef\vpen$$ 
where the $\delta_{n1}$ term is due to the appearance of 0 in just a single place in the anticonfined pattern, and where the factor 2 is due to the fact that a $0$ can arise from the confined pattern for $+1$ as well as from that for $-1$. We thus obtain the equation
$$U_n+U_{n-3}=2U_{n-1}+\delta_{n1}.\eqdef\vhex$$ 
The dynamical degree of the mapping, given by the largest root of the characteristic equation for (\vhex),  is precisely the golden mean already obtained by different methods above. 
The interesting point here is that, had we tried to compute the degree of the mapping from the number of spontaneous appearances of the value $\infty$, we would have found the obvious contribution $2U_{n-2}$ from the confined patterns, {\sl plus}  two contributions $f_{n-3}+\delta_{n2}$ and $f_{n}$ (where $f_n$ is defined by $f_{n+1}=f_n+f_{n-1}$ for $n\ge1$ with $f_1=1$ and $f_n=0$ for $n\le0$) due to the fact that $\infty$ appears an infinite number of times in the anticonfined pattern, with Fibonacci exponents. This would then result in an equation compatible with (\vhex), but with a source term that exhibits the same growth rate as that given by the homogeneous part of the equation, which of course does not change the value of the dynamical degree.
\bigskip
3. {\scap Conclusions}
\medskip
In this paper we have shown, through a selection of illustrative examples, how the method of Halburd, and the express variant of it we introduced in [\rodone] and [\rodnon], can be applied to the calculation of the dynamical degree of second-order rational mappings. The method is based on the singularity patterns of the mapping, using the information in them in order to establish linear equations that allow us to calculate the exact value of the dynamical degree.

The full method of Halburd allows for the exact calculation of the degree of the mapping, but necessitates the knowledge of all singularity patterns and not just the confined or unconfined ones. On the other hand, the express method, which does not use any information on cyclic (and, in general, anticonfined ones) does not actually yield the exact degree of each iterate but operates more like an integrability detector. Still, the express method allows for the exact calculation of the dynamical degree which is, after all, all that one needs to distinguish nonintegrable from integrable systems. One particularly interesting situation in this respect is the one  illustrated here by the mapping of Tsuda et al., namely the existence of an anticonfined singularity with exponential growth. In such case the growth in the order of the anticonfined singularity already constitutes a lower bound for the dynamical degree of the mapping itself, and in a general setting, the dynamical degree of such a mapping will be given by the fastest growth among the orders in the anticonfined patterns and the characteristic roots of the homogeneous equations obtained from the express method.

Linearisable mappings can also be accommodated within the present approach. The result of the calculation in this case is a dynamical degree equal to 1, independently of the nature (confined or not) of their singularities. 
\bigskip
{\scap Acknowledgements}
\medskip
TM and RW would like to acknowledge support from the Japan Society for the Promotion of Science (JSPS),  through the JSPS grants: KAKENHI grant number 16H06711 and KAKENHI grant number 15K04893. 
\bigskip
{\scap References}
\medskip
\item{[\sincon]} B. Grammaticos, A. Ramani and V. Papageorgiou, Phys. Rev. Lett. 67 (1991) 1825.
\item{[\arnold]} V. I. Arnold, Bol. Soc. Bras. Mat. 21 (1990) 1.
\item{[\veselov]} A.P. Veselov, Comm. Math. Phys. 145 (1992) 181.
\item{[\bellon]} M. Bellon and C-M. Viallet, Comm. Math. Phys. 204 (1999) 425.
\item{[\favre]} J. Diller and C. Favre, Amer. J. Math. 123 (2001) 1135.
\item{[\take]} T. Takenawa, J. Phys. A 34 (2001) 10533.
\item{[\rod]} R.G. Halburd, Proc. R. Soc. A 473 (2017) 20160831.
\item{[\anticonf]}  T. Mase, R. Willox, B. Grammaticos and A. Ramani, {\sl Integrable mappings and the notion of anticonfinement}, preprint (2017) arXiv:1511.02000v2 [math-ph].
\item{[\rodone]} A. Ramani, B. Grammaticos, R. Willox and T. Mase, J. Phys. A 50 (2017) 185203.
\item{[\rodnon]} A. Ramani, B. Grammaticos, R. Willox, T. Mase and J. Satsuma, {\sl Calculating the algebraic entropy of mappings with unconfined singularities}, preprint (2017) arXiv:1711.09237 [math-ph].
\item{[\redeem]} A. Ramani, B. Grammaticos, R. Willox, T. Mase and M. Kanki, J. Phys. A 48 (2015) 11FT02.
\item{[\hiv]} J. Hietarinta and C-M. Viallet, Phys. Rev. Lett. 81 (1998) 325.
\item{[\kim]} E. Bedford and K. Kim, Michigan Math. J. 54 (2006) 647.
\item{[\kanki]} M. Kanki, T. Mase and T. Tokihiro, J. Phys. A 48 (2015) 355202.
\item{[\linqrt]} A. Ramani, B. Grammaticos, J. Satsuma and N. Mimura, J. Phys. A 44 (2011) 425201.
\item{[\tsuda]}  T. Tsuda, A. Ramani, B. Grammaticos and T. Takenawa, Lett. Math. Phys. 82 (2007) 39.

\end